\documentclass[conference]{IEEEtran}
\usepackage[T1]{fontenc}
\usepackage[utf8]{inputenc}
\usepackage{amssymb}
\usepackage[pdftex]{graphicx}
\usepackage{subcaption}
\usepackage{amsmath}
\usepackage{amsthm}
\usepackage{mathtools}
\usepackage[noadjust]{cite}
\usepackage[acronym]{glossaries}
\usepackage{siunitx}
\usepackage{tikz}
\usepackage{flushend}
\usepackage{multirow}
\usepackage{pgfplots}
\usepackage{nicefrac}
\usepackage{fancyref}
\usepackage[ruled,boxed,lined,linesnumbered]{algorithm2e}
\usepackage[hidelinks]{hyperref}
\usepackage[capitalise]{cleveref}
\usepackage{lipsum}
\pgfplotsset{compat=1.17}

\usetikzlibrary{arrows,chains,matrix,positioning,scopes}
\usetikzlibrary{calc}
\usetikzlibrary{fit,shapes}

\newacronym{NPRACH}{NPRACH}{narrowband physical random-access channel}
\newacronym{ToA}{ToA}{time of arrival}
\newacronym{CFO}{CFO}{carrier frequency offset}
\newacronym{NBIoT}{NB-IoT}{narrowband internet of things}
\newacronym{5GNR}{5G NR}{5G New Radio}
\newacronym{3GPP}{3GPP}{3rd Generation Partnership Project}
\newacronym{UMi}{UMi}{urban microcell}
\newacronym{RMSE}{RMSE}{root-mean-square error}
\newacronym{NN}{NN}{neural network}
\newacronym{BS}{BS}{base station}
\newacronym{UE}{UE}{user equipment}
\newacronym{SG}{SG}{symbol group}
\newacronym{CP}{CP}{cyclic prefix}
\newacronym{OFDM}{OFDM}{orthogonal frequency division multiplexing}
\newacronym{FFT}{FFT}{fast Fourier transform}
\newacronym{AWGN}{AWGN}{additive white Gaussian noise}
\newacronym{DFT}{DFT}{discrete Fourier transform}
\newacronym{FNR}{FNR}{false negative rate}
\newacronym{FPR}{FPR}{false positive rate}
\newacronym{RG}{RG}{resource grid}
\newacronym{RE}{RE}{resource element}
\newacronym{SNR}{SNR}{signal-to-noise ratio}
\newacronym{1D}{1D}{one-dimensional}
\newacronym{MLP}{MLP}{multilayer perceptron}
\newacronym{BCE}{BCE}{binary cross-entropy}
\newacronym{KL}{KL}{Kullback–Leibler}
\newacronym{SGD}{SGD}{stochastic gradient descent}
\newacronym{ppm}{ppm}{parts-per-million}
\newacronym{ICI}{ICI}{inter-carrier interference}
\newacronym{GNN}{GNN}{graph neural network}
\newacronym{BP}{BP}{belief propagation}
\newacronym{FEC}{FEC}{forward error correction}
\newacronym{LDPC}{LDPC}{low-density parity-check}
\newacronym{HDPC}{HDPC}{high-density parity-check}
\newacronym{SCL}{SCL}{successive cancellation list}
\newacronym{SC}{SC}{successive cancellation}
\newacronym{URLLC}{URLLC}{ultra-reliable low-latency communications}
\newacronym{APP}{APP}{a posterior probability}
\newacronym{MIMO}{MIMO}{multiple-input multiple-output}
\newacronym{CNN}{CNN}{convolutional neural network}
\newacronym{BER}{BER}{bit error rate}
\newacronym{BPSK}{BPSK}{binary phase shift keying}
\newacronym{LLR}{LLR}{log-likelihood ratio}
\newacronym{FN}{FN}{factor node}
\newacronym{VN}{VN}{variable node}
\newacronym{CN}{CN}{check node}
\newacronym{MPNN}{MPNN}{message passing neural network}

\newacronym{AI}{AI}{artificial intelligence}
\newacronym{ML}{ML}{machine learning}
\newacronym{SISO}{SISO}{single input single output}
\newacronym{PRB}{PRB}{physical resource block}
\newacronym{PUSCH}{PUSCH}{physical uplink shared channel}

\newacronym{MUMIMO}{MU-MIMO}{multi-user multiple-input multiple-output}
\newacronym{BICM}{BICM}{bit-interleaved coded modulation}
\newacronym{QAM}{QAM}{quadrature amplitude modulation}
\newacronym{LMMSE}{LMMSE}{linear minimum mean square error}
\newacronym{CSI}{CSI}{channel state information}
\newacronym{SIMO}{SIMO}{single-input multiple-output}
\newacronym{CGNN}{CGNN}{convolutional and graph neural network}
\newacronym{BLER}{BLER}{block error rate}
\newacronym{LS}{LS}{least squares}
\newacronym{PE}{PE}{positional encoding}
\newacronym{relu}{ReLU}{rectified linear unit}
\newacronym{RB}{RB}{resource block}
\newacronym{CGGNN}{CGGNN}{convolutional graph neural network}
\newacronym{DMRS}{DMRS}{demodulation reference signal}
\newacronym{IoT}{IoT}{internet of things}
\newacronym{ADAM}{ADAM}{adaptive momentum}
\newacronym{TBLER}{TBLER}{transport block error rate}
\newacronym{MCS}{MCS}{modulation and coding scheme}
\newacronym{TDL}{TDL}{tapped delay line}
\newacronym{CDM}{CDM}{code division multiplexing}
\newacronym{FLOP}{FLOP}{floating point operation}
\newacronym{PHY}{PHY}{physical layer}

\newacronym{ULA}{ULA}{uniform linear array}
\newacronym{NRX}{NRX}{neural receiver}
\newacronym{Var-MCS-NRX}{Var-MCS NRX}{variable-MCS NRX}
\newacronym{UL}{UL}{uplink}
\newacronym{DL}{DL}{downlink}
\newacronym{MSE}{MSE}{mean squared error}
\newacronym{CIR}{CIR}{channel impulse response}
\newacronym{iid}{iid}{independent and identically distributed}

\renewcommand{\vec}[1]{\mathbf{#1}}
\newcommand{\vecs}[1]{\boldsymbol{#1}}

\newcommand{\nv}{\vec{n}}

\newcommand{\vv}{\vec{v}}

\newcommand{\xv}{\vec{x}}
\newcommand{\yv}{\vec{y}}

\newcommand{\ellv}{\vecs{\ell}}

\newcommand{\Hm}{\vec{H}}
\newcommand{\Id}{\vec{I}}

\newcommand{\Sm}{\vec{S}}

\newcommand{\CC}{\mathbb{C}}

\newcommand{\RR}{\mathbb{R}}

\newcommand{\LB}{\left(}
\newcommand{\RB}{\right)}

\renewcommand{\ln}[1]{\mathop{\mathrm{ln}}\LB #1\RB}

\definecolor{mittelblau}{RGB}{0, 126, 198}
\definecolor{violettblau}{cmyk}{0.9, 0.6, 0, 0}
\definecolor{rot}{RGB}{238, 28 35}
\definecolor{apfelgruen}{RGB}{140, 198, 62}
\definecolor{gelb}{RGB}{1, 221, 0}
\definecolor{orange}{RGB}{244, 111, 33}
\definecolor{pink}{RGB}{237, 0, 140}
\definecolor{lila}{RGB}{128, 10, 145}
\definecolor{hellgrau}{RGB}{224, 224, 224}
\definecolor{mittelgrau}{RGB}{128, 128, 128}
\definecolor{dunkelgrau}{RGB}{80,80,80}
\definecolor{anthrazit}{RGB}{19, 31, 31}

\pgfplotscreateplotcyclelist{corporate_colours_markers}{%
rot, every mark/.append style={fill=.!80!rot},mark=*\\%
mittelblau, every mark/.append style={fill=.!80!mittelblau},mark=square*\\%
apfelgruen, every mark/.append style={fill=.!80!apfelgruen},mark=triangle*\\%
violettblau, every mark/.append style={fill=.!80!violettblau},mark=otimes*, densely dotted\\%
orange, mark=star, densely dotted\\%
pink, every mark/.append style={fill=.!80!pink},mark=diamond*, densely dashed\\%
lila, mark=|, densely dashed\\%
gelb, every mark/.append style={fill=.!80!gelb},mark=pentagon*\\%
hellgrau, mark=text,text mark=p\\%
anthrazit, mark=text,text mark=a\\%
}

\pgfplotscreateplotcyclelist{corporate_colours_no_markers}{%
rot, every mark/.append style={fill=.!80!rot},mark=none\\%
mittelblau, every mark/.append style={fill=.!80!mittelblau},mark=none\\%
apfelgruen, every mark/.append style={fill=.!80!apfelgruen},mark=none\\%
violettblau, every mark/.append style={fill=.!80!violettblau},mark=none, densely dotted\\%
orange, mark=none, densely dotted\\%
pink, every mark/.append style={fill=.!80!pink},mark=none, densely dashed\\%
lila, mark=none, densely dashed\\%
gelb, every mark/.append style={fill=.!80!gelb},mark=none\\%
hellgrau, mark=text,text mark=none\\%
anthrazit, mark=text,text mark=none\\%
}

\pgfplotscreateplotcyclelist{corporate_colours_no_markers1}{%
anthrazit, mark=none\\%
rot, every mark/.append style={fill=.!80!rot},mark=none\\%
violettblau, every mark/.append style={fill=.!80!violettblau},mark=none\\%
apfelgruen, every mark/.append style={fill=.!80!apfelgruen},mark=none\\%
orange, mark=none\\%
pink, every mark/.append style={fill=.!80!pink},mark=none\\%
mittelblau, every mark/.append style={fill=.!80!mittelblau},mark=none\\%
lila, mark=none, densely dashed\\%
gelb, every mark/.append style={fill=.!80!gelb},mark=none\\%
gray, mark=text,text mark=none\\%
}

\pgfplotscreateplotcyclelist{corporate colours markers double}{%
rot, every mark/.append style={fill=.!80!rot},mark=*\\%
rot, dashed, every mark/.append style={fill=.!80!rot,solid},mark=*\\%
mittelblau, every mark/.append style={fill=.!80!mittelblau},mark=square*\\%
mittelblau, dashed, every mark/.append style={fill=.!80!mittelblau,solid},mark=square*\\%
apfelgruen, every mark/.append style={fill=.!80!apfelgruen},mark=triangle*\\%
apfelgruen, dashed, every mark/.append style={fill=.!80!apfelgruen,solid},mark=triangle*\\%
orange, mark=star\\%
orange, dashed, mark=star\\%
pink, every mark/.append style={fill=.!80!pink},mark=diamond*\\%
pink, dashed, every mark/.append style={fill=.!80!pink,solid},mark=diamond*\\%
violettblau, every mark/.append style={fill=.!80!violettblau},mark=otimes*\\%
violettblau, dashed, every mark/.append style={fill=.!80!violettblau,solid},mark=otimes*\\%
lila, mark=|\\%
lila, dashed, mark=|\\%
gelb, every mark/.append style={fill=.!80!gelb},mark=pentagon*\\%
gelb, dashed, every mark/.append style={fill=.!80!gelb,solid},mark=pentagon*\\%
}

\IEEEoverridecommandlockouts

\newcommand\comsc[1]{\marginpar{\textcolor{rot}{\small{#1}}}}

\begin{document}

\newpage
\setcounter{page}{0}

\title{Design of a Standard-Compliant Real-Time \newline Neural Receiver for 5G NR}

\author{
\IEEEauthorblockN{Reinhard Wiesmayr$^{\S,\ast}$, Sebastian Cammerer$^\dagger$, Fay\c{c}al A\"{i}t Aoudia$^\dagger$, Jakob Hoydis$^\dagger$}
\IEEEauthorblockN{Jakub Zakrzewski$^\dagger$, and Alexander Keller$^\dagger$}

\IEEEauthorblockA{$^\dagger$NVIDIA, $^\S$ETH Zurich, contact: scammerer@nvidia.com}

\thanks{$^\ast$Work done during an internship at NVIDIA.}

\thanks{This work has received financial support from the European Union under Grant Agreement 101096379 (CENTRIC). Views and opinions expressed are however those of the author(s) only and do not necessarily reflect those of the European Union or the European Commission (granting authority). Neither the European Union nor the granting authority can be held responsible for~them.}
\vspace*{-0.5cm}
}

\maketitle

\glsresetall


\begin{abstract}

We detail the steps required to deploy a \gls{MUMIMO} \gls{NRX} in an actual cellular communication system. This raises several exciting research challenges, including the need for real-time inference and compatibility with the 5G NR standard.
As the network configuration in a practical setup can change dynamically within milliseconds, we propose an adaptive \gls{NRX} architecture capable of supporting dynamic \gls{MCS} configurations without the need for any re-training and without additional inference cost.
We optimize the latency of the \gls{NN} architecture to achieve inference times of less than 1ms on an NVIDIA A100 GPU using the TensorRT inference library. These latency constraints effectively limit the size of the \gls{NN} and we quantify the resulting \gls{SNR} degradation as less than 0.7\,dB when compared to a preliminary non-real-time \gls{NRX} architecture.
Finally, we explore the potential for site-specific adaptation of the receiver by investigating the required size of the training dataset and the number of fine-tuning iterations to optimize the \gls{NRX} for specific radio environments using a ray tracing-based channel model.
The resulting \gls{NRX} is ready for deployment in a real-time 5G NR system and the source code including the TensorRT experiments is available online.\footnote{\url{https://github.com/nvlabs/neural_rx}}

\end{abstract}

\glsresetall


\section{Introduction}

Significant performance gains have been demonstrated by \gls{NN}-based signal processing for wireless communications \cite{o2017introduction, honkala2021deeprx,aoudia2021end,ye2017power, lin2023artificial}. However, little has been reported regarding the practical deployment of these algorithms in a 5G NR system.
The challenges include a) real-time inference imposing  strict latency constraints on the \gls{NN} architecture, b) the need for a dynamic re-configuration of the \gls{MCS} without re-training, and c) site-specific fine-tuning to adapt the algorithms to the specific environment which even allows for continuous performance improvement after deployment.

In this work, we focus on the concept of \glspl{NRX} \cite{honkala2021deeprx,ye2017power,cammerer2023neural}, where a single \gls{NN} is trained to jointly perform channel estimation, equalization, and demapping. The concept has been first introduced in \cite{honkala2021deeprx} with \gls{MIMO} extensions in \cite{korpi2021deeprx}. A 5G NR compliant multi-user \gls{MIMO} receiver for the \gls{PUSCH} is proposed in \cite{cammerer2023neural}. We revise our  architecture from \cite{cammerer2023neural} and investigate two methods to support dynamic \gls{MCS} configurations without the need for re-training.

Predicting the inference latency of a given neural network architecture is a challenging task for which the results strongly depend on the targeted hardware platform, the specific software stack as well as the level of code optimization.
Thus, the number of \glspl{FLOP}, weights, or layers is often used as surrogate metric to predict the model's computational complexity. However, such metrics may lead to inaccurate conclusions due to the high level of parallelism and unknown memory bottlenecks during inference.
We deploy our \gls{NRX} \cite{cammerer2023neural} using the TensorRT inference library on the targeted NVIDIA A100 GPU platform. This ensures realistic latency measurements and allows for eliminating bottlenecks from the critical path.
As a result, we propose a carefully optimized real-time version of the \gls{NRX} architecture.

Another interesting aspect of \gls{ML}-based receiver design is the inherent possibility of data-aided fine-tuning and site-specific adaptation of the algorithms. Contrary to (accidental) overfitting of the \gls{NN}, the idea is to let the receiver learn the underlying channel statistics of the specific deployment scenario. Early promising results have been reported in \cite{fischer2022adaptive1} for a single user system, and recently in \cite{Uzlaner2024dynamic}.

However, it remains a challenge to gather a sufficiently diverse real-world dataset to validate the possible gains and to quantify the required amount of training data. Some static users may produce many similar channel realizations, while a few dynamic users may generate a much richer dataset.
We use ray tracing \cite{hoydis2023sionna} to generate environment-specific \glspl{CIR} and investigate the required number of samples and training iterations for \gls{NRX} fine-tuning.

Given the strict real-time constraints, we will address a similar research question as in \cite{fischer2022adaptive1} of whether a large neural network trained offline to generalize to arbitrary channel conditions performs better than a smaller, adaptive network optimized for specific scenarios. As it turns out, this principle of \emph{generalization through adaptability} offers a competitive performance under strictly limited computational complexity and latency, though it may require additional training resources and may reduce robustness to unforeseen conditions. We like to emphasize that we do \emph{not} assume training \emph{on-the-fly} (i.e., online). Instead, we train periodically (or even just once) on a small dataset which can be regarded a by-product of the normal operation mode of the receiver.


\section{Background}

We assume \gls{MUMIMO}\glsadd{MIMO} \gls{UL} transmission from $U$ \glspl{UE} to a single \gls{BS} with $B$ antennas.
While each \gls{UE} can have multiple transmit antennas (e.g., \gls{UE} $u$ is equipped with $N_u$ antennas), we assume that each \gls{UE} only transmits a single \gls{MIMO} stream.\footnote{In 5G NR terminology, a \gls{MIMO} stream is often called \emph{layer}.}

We consider transmission over the 5G NR \gls{PUSCH} adhering to a standard-compliant \gls{OFDM} frame structure with $S$ subcarriers and $T=14$ \gls{OFDM} time symbols per slot\footnote{The preliminary \gls{NRX} architecture from \cite{cammerer2023neural} as well as our extensions adapt to varying values of $T$ and $S$ without the need for retraining.} and refer the interested reader to \cite{dahlman20205g} for more details on 5G NR systems.
With a sufficiently long cyclic prefix, the \gls{MUMIMO} input-output relation on each subcarrier $s\in\{1,\dots, S\}$ and for each \gls{OFDM} time symbol $t\in\{1,\dots, T\}$ can be modeled as
\begin{equation}
    \yv_{s,t} = \sum_{u=1}^U \Hm_{s,t,u} \xv_{s,t,u} + \nv_{s,t}
\end{equation}
where $\yv_{s,t}\in\CC^{B}$ is the received signal, $\Hm_{s,t,u}\in\CC^{B \times N_u}$ is the \gls{MIMO} channel matrix, $\xv_{s,t,u}$ is the modulated transmit vector of \gls{UE} $u$ after beamforming, and $\nv_{s,t}\sim \mathcal{NC}(0, N_0 \Id)$ is the complex Gaussian noise  with power spectral density $N_0$.

While most of the \glspl{RE} (indexed by the tuple $(s,t)$) are allocated for data transmission, certain \glspl{RE} are reserved for pilot symbols (called \gls{DMRS}) which are known to the \gls{BS} and used for channel estimation.
For simplicity, we will omit the indices $s$ and $t$ in the following.

The \glsplural{UE} apply codebook-based beamforming $\xv_u = \vv_u \tilde{x}_u$ where $\vv_u$ is a beamforming vector and $\tilde{x}_u$ is the modulated transmit symbol which is taken from an $2^m$-ary constellation. Throughout this paper, we focus on \gls{MCS} indices $i$ from 5G NR described in \cite[Table 5.1.3.1-1]{38214}, which applies QPSK, 16-QAM, and 64-QAM, with varying code rates.
The $m$ bits transmitted in $\tilde{x}_u$ originate from random payload bits that are encoded by 5G NR compliant \gls{LDPC} channel coding and rate-matching, which depend on the \gls{MCS} index $i$ and the total number of data-carrying \glspl{RE}.

The goal of classical \gls{MIMO} detectors as well as that of the \gls{NRX} is to compute \gls{LLR} estimates for each of the \gls{UE}'s transmitted bits $b$ from the received signal.
We define the \glspl{LLR} as logits, i.e.,
\begin{equation}
    \ell = \ln{\frac{\Pr(b=1|\yv)}{\Pr(b=0|\yv)}} \label{eq:llr}
\end{equation}
and feed their estimates to the subsequent channel decoder.

\subsection{Preliminary Neural Receiver Architecture}

\begin{figure}[t]
    \centering
    \includegraphics[width=.95\columnwidth]{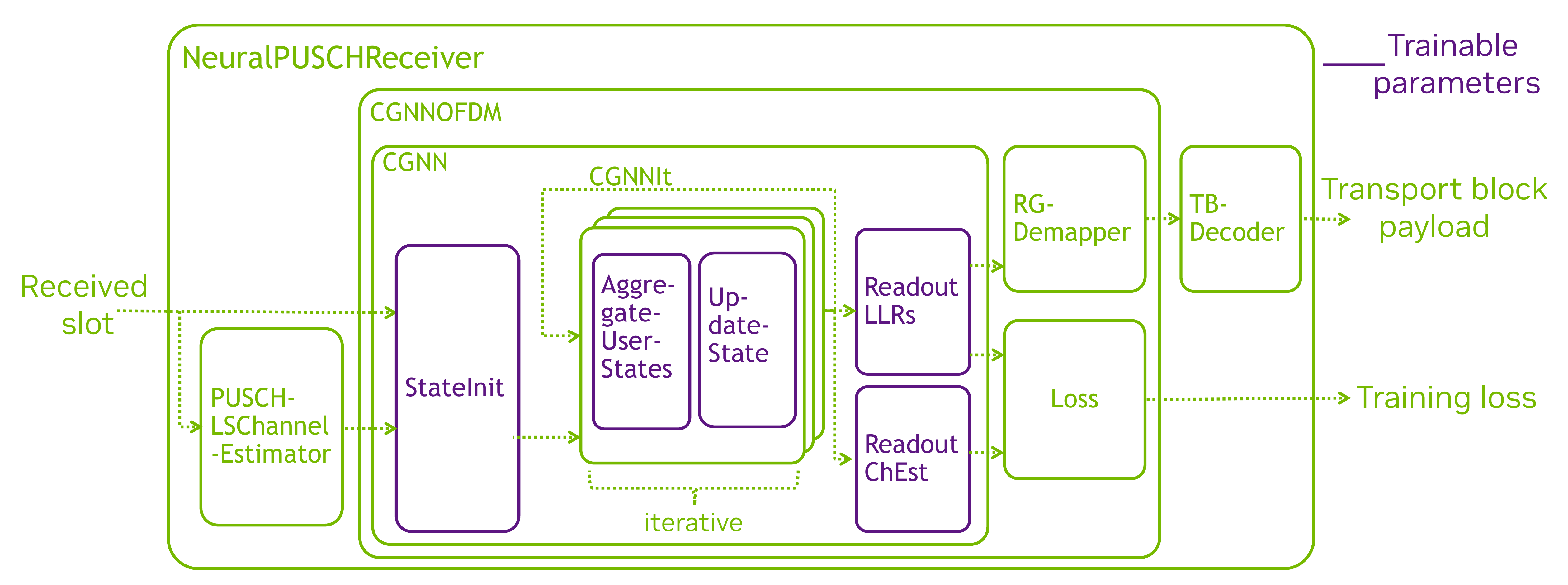}
    \vspace*{-0.2cm}
    \caption{\Glsdesc{NRX} for 5G NR \gls{PUSCH}.}
    \label{fig:nrx_architecture}
\end{figure}

We briefly revisit the \gls{NRX} architecture from \cite{cammerer2023neural} that was proposed for a single \gls{MCS} and which we will extend to support varying and mixed \gls{MCS} in \cref{sec:var_mcs}.

The \gls{NRX} depicted in \cref{fig:nrx_architecture} implements a \gls{CGNN} for \gls{MUMIMO} detection across the \gls{OFDM} \gls{RG}. To enable 5G NR standard compliance, the \gls{CGNN} is surrounded by a \gls{RG} demapper and a transport block decoder (cf. Sionna's 5G NR module\footnote{\url{https://nvlabs.github.io/sionna/api/nr.html}} for details). To enable channel estimation for varying \gls{DMRS}, e.g., resulting from varying slot or \gls{UE} indices, a \gls{LS} channel estimator provides an initial channel estimate to the \gls{CGNN} for each individual data stream.

The \gls{CGNN} architecture consists of three main components: (i) the state initialization layer (\texttt{StateInit}), (ii) the unrolled iterative \gls{CGNN} algorithm (\texttt{CGNNit} blocks), and (iii) the read-out layers. While the architecture in \cite{cammerer2023neural} only proposes one type of output layer that is used to transform the state variable to LLR estimates (\texttt{ReadoutLLRs}), our new architecture implements an additional read-out layer that outputs a (refined) channel estimate, also from the same state variable.

As detailed in \cite{cammerer2023neural}, the \texttt{StateInit} layer implements a small \gls{CNN} that transforms the \gls{CGNN} inputs, i.e., the entire \gls{RG} of received signals $\yv_{s,t}$, a positional encoding of distances to the next pilot symbol, and each \gls{UE}'s initial \gls{LS} channel estimates, into the initial state vector $\Sm_u^{(0)}\in\RR^{S\times T \times d_{\textnormal{S}}}$ of depth $d_{\textnormal{S}}$. The main part of the \gls{NRX} consists of an unrolled iterative algorithm that applies $N_{\textnormal{it}}$ consecutive \texttt{CGNNit} blocks, each of which updates the state vectors of all \glspl{UE} in parallel. Each \texttt{CGNNit} block first performs a message passing step where a \gls{MLP} is used to transform each \gls{UE}'s state to messages, which are then aggregated by taking for each \gls{UE} the sum of messages from all other \glspl{UE}. The second part of each \texttt{CGNNit} block is composed of a \gls{CNN} that updates each \gls{UE}'s state based on the previous iteration's state, the aggregated message-passing messages, and the positional encoding.
After $N_{\textnormal{it}}$ such \texttt{CGNNit} blocks, readout \glspl{MLP} are applied to transform the final state vectors $\Sm_u^{(N_{\textnormal{it}})}$ into the desired outputs. As in \cite{cammerer2023neural}, the \texttt{ReadoutLLRs} layer outputs \gls{LLR} estimates $\hat\ellv_{s,t,u}\in\RR^{m}$ for each \gls{UE}'s symbol of all \glspl{RE} on the \gls{RG}.

We extend the \gls{NRX} from \cite{cammerer2023neural} by an additional read-out layer, denoted as \texttt{ReadoutChEst}, that outputs refined channel estimates computed from the same state vectors $\Sm_u^{(N_{\textnormal{it}})}$.
This \texttt{ReadoutChEst} layer not only provides more accurate channel estimates, it is also found to improve the training convergence when applied using an  additional \gls{MSE} loss term as described in the following.

\subsection{Training Scheme of the Neural Receiver}\label{sec:preliminary_training}

The \gls{NRX} from \cite{cammerer2023neural} is trained by empirical risk minimization with the \gls{BCE} loss that is computed between the \gls{LLR} estimates and ground-truth bit labels. A training step describes one gradient-based weight update that is computed from the average loss of $N$ independent samples. Each of these $N$ batch samples represents transmission of one entire \gls{OFDM} \gls{RG}. If not stated otherwise, the \gls{NRX} is trained on synthetic training data sampled from the 3GPP \gls{UMi} channel model.
For each batch, the number of active users $1\leq U_{\textnormal{A}} \leq U$ is randomly sampled from a triangular distribution \cite{9298921}, and, in each batch sample, each \gls{UE} transmits random payload bits that are individually encoded and modulated. The \gls{SNR} is random-uniformly sampled (in the log-scale) for each batch sample from a pre-defined \gls{SNR} range, which is a hyperparameter.

We propose the following extensions to the training scheme from \cite{cammerer2023neural} and detail additional considerations for \gls{Var-MCS-NRX} in \cref{sec:var_mcs_training}.
\subsubsection{Double-readout}
The \texttt{ReadoutChEst} layer can be jointly trained with the \texttt{ReadoutLLRs} layer by adding an \gls{MSE} loss to the \gls{BCE} loss.
The \gls{MSE} loss is computed between the output channel estimates and the ground-truth channel realizations, and scaled by a hyperparameter $\gamma$ that controls its contribution to the total loss (i.e., the sum of \gls{BCE} and scaled \gls{MSE} loss) used for gradient computation.

\subsubsection{Multi-loss}
To support a variable number of unrolled \gls{NRX} iterations $1 \leq N_{\textnormal{it}}^\prime \leq N_{\textnormal{it}}$, the \gls{NRX} is trained with the so-called multi-loss \cite{nachmani2016learning}. There, the read-out layers are applied to the state variable after each \gls{NRX} iteration and the total loss is accumulated from the loss of all $N_{\textnormal{it}}$ model readouts.


\section{Neural Receivers with Variable MCS}\label{sec:var_mcs}

In this paper, we study two approaches to extend the \gls{NRX} architecture from \cite{cammerer2023neural} to support variable modulation schemes\footnote{The code rate and coding scheme is transparent to the \gls{NRX}, as the \gls{NRX} outputs \glspl{LLR} on coded bits.}: (i) \emph{masking} of higher-order \glspl{LLR}, and (ii) \emph{\gls{MCS}-specific input and output layers} (abbreviated by \emph{Var-IO}).

The first method builds upon an idea mentioned in \cite{honkala2021deeprx}, which was later implemented for a neural demapper in \cite{gansekoele2024machine}. The working principle builds upon the recursive structure of bit labels from Gray-code-labeled QAM constellations, where constellation points from higher-order modulations are recursively derived from lower-order points. The higher-order bit labels then re-apply the lower-order bit labels, and are extended by the additional higher-order bits. For example, if we compare 16-QAM to QPSK, all 16-QAM constellation points within a quadrant of the complex plane have lower-order bits identical to the corresponding QPSK constellation point.
Thus, by masking of unused higher-order LLR outputs, an \gls{NRX} for the highest modulation order can be applied for detecting lower-order constellation points, too. As mentioned in \cref{sec:var_mcs_training}, training such a \gls{Var-MCS-NRX} requires additional considerations.

Note that masking can be also applied to classical \gls{LLR} demapping algorithms, e.g., to \gls{APP} demapping. Though such \emph{mismatched demapping} will produce \glspl{LLR} with the same sign as a \emph{matched demapper} for the correct constellation,
the \gls{LLR} magnitudes do in general not match the underlying probabilities, as defined in \eqref{eq:llr}.
Without additional \gls{LLR} correction, such as proposed in \cite{studer2010soft}, ``classical'' mismatched demapping can lead to severe performance degradation in \gls{FEC} decoding.

\begin{figure}[t]
    \centering
    \includegraphics[width=\columnwidth]{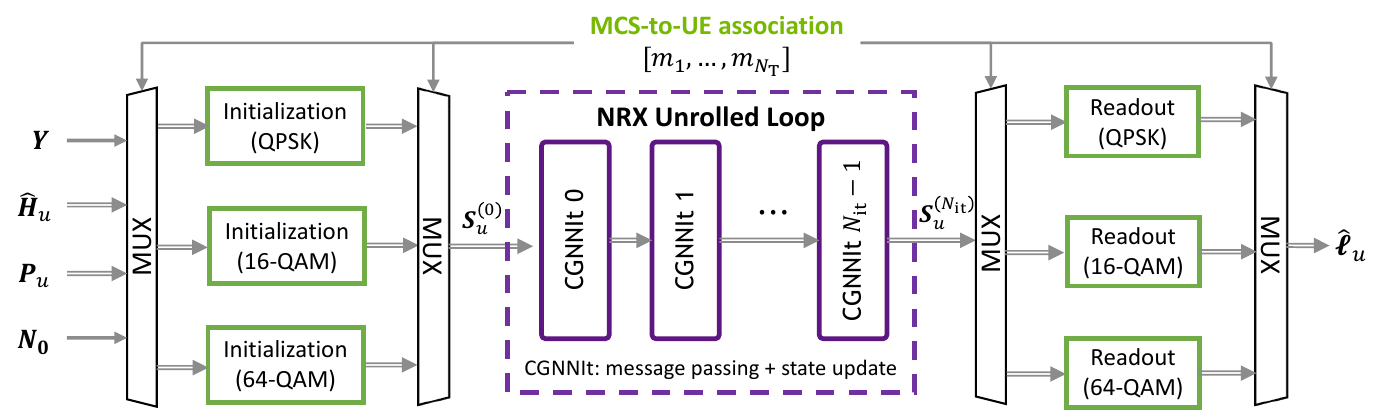}
    \caption{\gls{Var-MCS-NRX} architecture with Var-IO layers.}
    \label{fig:var-mcs-var-io}
\end{figure}

While our proposed training scheme discussed in \cref{sec:var_mcs_training} turned out effective to train a \gls{Var-MCS-NRX} with the masking scheme \cite{gansekoele2024machine} for Gray-code labeled QAM constellations, we also put forward an alternative method for implementing the \gls{Var-MCS-NRX}. By applying modulation-specific \texttt{StateInit} and \texttt{ReadoutLLRs} layers (denoted as input and output layers, respectively), we allow the \gls{NRX} to fit to varying modulation orders, while sharing the majority of weights (in the \texttt{CGNIt} blocks) across modulation schemes. Modulation-specific input layers are motivated by improving data-aided channel estimation. The intuition for modulation-specific output layers is to enable the model to learn \emph{matched} demapping. Only the number of \gls{LLR} outputs needed for the corresponding modulation order has to be implemented. This Var-IO scheme (as depicted in \cref{fig:var-mcs-var-io}) can be also useful with more general, non-Gray-code-labeled constellations, or custom constellations obtained from end-to-end learning \cite{o2017introduction}. Note that although \gls{MCS}-specific IO layers lead to a slightly increased number of weights, the number of active weights (and, thus, the inference latency) is the same as for the single-\gls{MCS} \gls{NRX}.

\begin{figure*}[ht]
    \centering
    \begin{tikzpicture}
        \pgfplotsset{compat=1.5}
        \tikzset{font={\fontsize{7pt}{7}\selectfont}}
        \begin{axis}[
            hide axis,
            width=.9\textwidth,
            height=2cm,
            line width=1pt,
            xmin=0.9,
            xmax=1.0,
            ymin=0.9,
            ymax=1.0,
            legend style={draw=white!15!black,legend cell align=left,nodes={scale=0.945, transform shape}},
            legend columns=7,
            cycle list name=corporate_colours_no_markers]
            \pgfplotsinvokeforeach{1,...,7}{\addplot coordinates {(0,0)};}
            \addlegendentry{LS + LMMSE};
            \addlegendentry{LMMSE + K-Best};
            \addlegendentry{Perf.-CSI + K-Best};
            \addlegendentry{RT NRX};
            \addlegendentry{Large NRX};
            \addlegendentry{RT Var-MCS NRX Var-IO};
            \addlegendentry{Large Var-MCS NRX Mask.};
        \end{axis}
    \end{tikzpicture}
    \begin{subfigure}{0.32\textwidth}
        \centering
    \begin{tikzpicture}
        \pgfplotsset{compat=1.5}
        \tikzset{font={\fontsize{7pt}{7}\selectfont}}
        \begin{axis}[
            xmode=normal,
            ymode=log,
            xlabel={SNR [dB]},
            ylabel=$\mathrm{TBLER}$,
            y label style={at={(axis description cs:-0.15,.5)},anchor=south},
            x label style={at={(axis description cs:0.5,-0.075)},anchor=north},
            xmin = -3,
            xmax = 5,
            ymax = 1e-0,
            ymin = 7e-4,
            mark size=1.5pt,
            legend style={nodes={scale=0.68, transform shape}},
            legend pos = north east,
            grid=both,
            minor grid style={gray!25},
            major grid style={gray!25},
            width=1.1\textwidth,
            height=4.5cm,
            legend cell align={left},
            line width=1pt,
            cycle list name=corporate_colours_no_markers]
            \addplot table[x=snr_db, y=BaselineLSlinLMMSE20, col sep=comma]{./results/csv/2024-07-18_17-55_nrx_rt_qpsk.csv};
            \addplot table[x=snr_db, y=BaselineLMMSEKBest20, col sep=comma]{./results/csv/2024-07-18_17-55_nrx_rt_qpsk.csv};
            \addplot table[x=snr_db, y=BaselinePerfCSIKBest20, col sep=comma]{./results/csv/2024-07-18_17-55_nrx_rt_qpsk.csv};
            \pgfplotsset{cycle list shift=1}
            \addplot table[x=snr_db, y=NeuralReceiver20, col sep=comma]{./results/csv/2024-07-18_17-55_nrx_large_qpsk.csv};
            \addplot table[x=snr_db, y=NeuralReceiver20, col sep=comma]{./results/csv/2024-08-20_13-40_nrx_rt_var_mcs.csv};
            \addplot table[x=snr_db, y=NeuralReceiver20, col sep=comma]{./results/csv/2024-07-23_12-50_nrx_large_var_mcs_64qam_masking.csv};
        \end{axis}
    \end{tikzpicture}
    \caption{QPSK}
    \end{subfigure}
    \hfill
    \begin{subfigure}{0.32\textwidth}
    \begin{tikzpicture}
        \pgfplotsset{compat=1.5}
        \tikzset{font={\fontsize{7pt}{7}\selectfont}}
        \begin{axis}[
            xmode=normal,
            ymode=log,
            xlabel={SNR [dB]},
            y label style={at={(axis description cs:-0.1,.5)},anchor=south},
            x label style={at={(axis description cs:0.5,-0.075)},anchor=north},
            xmin = -1,
            xmax = 6,
            ymax = 1e-0,
            ymin = 7e-4,
            mark size=1.5pt,
            legend style={nodes={scale=0.85, transform shape}},
            legend pos = south west,
            grid=both,
            minor grid style={gray!25},
            major grid style={gray!25},
            width=1.1\textwidth,
            height=4.5cm,
            legend cell align={left},
            line width=1pt,
            cycle list name=corporate_colours_no_markers]
            \addplot table[x=snr_db, y=BaselineLSlinLMMSE20, col sep=comma]{./results/csv/2024-07-23_12-50_nrx_rt.csv};
            \addplot table[x=snr_db, y=BaselineLMMSEKBest20, col sep=comma]{./results/csv/2024-07-23_12-50_nrx_rt.csv};
            \addplot table[x=snr_db, y=BaselinePerfCSIKBest20, col sep=comma]{./results/csv/2024-07-23_12-50_nrx_rt.csv};
            \addplot table[x=snr_db, y=NeuralReceiver20, col sep=comma]{./results/csv/2024-07-23_12-50_nrx_rt.csv};
            \addplot table[x=snr_db, y=NeuralReceiver20, col sep=comma]{./results/csv/2024-07-18_17-55_nrx_large.csv};
            \addplot table[x=snr_db, y=NeuralReceiver21, col sep=comma]{./results/csv/2024-08-20_13-40_nrx_rt_var_mcs.csv};
            \addplot table[x=snr_db, y=NeuralReceiver21, col sep=comma]{./results/csv/2024-07-23_12-50_nrx_large_var_mcs_64qam_masking.csv};
        \end{axis}
    \end{tikzpicture}
    \caption{16-QAM}
    \end{subfigure}
    \hfill
    \begin{subfigure}{0.32\textwidth}
    \begin{tikzpicture}
        \pgfplotsset{compat=1.5}
        \tikzset{font={\fontsize{7pt}{7}\selectfont}}
        \begin{axis}[
            xmode=normal,
            ymode=log,
            xlabel={SNR [dB]},
            y label style={at={(axis description cs:-0.1,.5)},anchor=south},
            x label style={at={(axis description cs:0.5,-0.075)},anchor=north},
            xmin = 2,
            xmax = 9,
            ymax = 1e-0,
            ymin = 7e-4,
            mark size=1.5pt,
            legend style={nodes={scale=0.85, transform shape}},
            legend pos = south west,
            grid=both,
            minor grid style={gray!25},
            major grid style={gray!25},
            width=1.1\textwidth,
            height=4.5cm,
            legend cell align={left},
            line width=1pt,
            cycle list name=corporate_colours_no_markers]
            \addplot table[x=snr_db, y=BaselineLSlinLMMSE20, col sep=comma]{./results/csv/2024-07-23_12-50_nrx_large_64qam.csv};
            \addplot table[x=snr_db, y=BaselineLMMSEKBest20, col sep=comma]{./results/csv/2024-07-23_12-50_nrx_large_64qam.csv};
            \addplot table[x=snr_db, y=BaselinePerfCSIKBest20, col sep=comma]{./results/csv/2024-07-23_12-50_nrx_large_64qam.csv};
            \pgfplotsset{cycle list shift=1}
            \addplot table[x=snr_db, y=NeuralReceiver20, col sep=comma]{./results/csv/2024-07-23_12-50_nrx_large_64qam.csv};
            \pgfplotsset{cycle list shift=2}
            \addplot table[x=snr_db, y=NeuralReceiver22, col sep=comma]{./results/csv/2024-07-23_12-50_nrx_large_var_mcs_64qam_masking.csv};
        \end{axis}
    \end{tikzpicture}
    \caption{64-QAM}
    \end{subfigure}
    \vspace*{-0.1cm}
    \caption{\Gls{TBLER} vs. \gls{SNR} for a Double-TDL channel with $U \times B \equiv 2\times4$. The MCS indices are $i=9$ (left), $i=14$ (middle) and $i=19$ (right). Dotted curves denote single-MCS NRXs and the Var-MCS NRXs are dashed.}
    \label{fig:variable_mcs_big_system}
    \vspace*{-0.6cm}
\end{figure*}
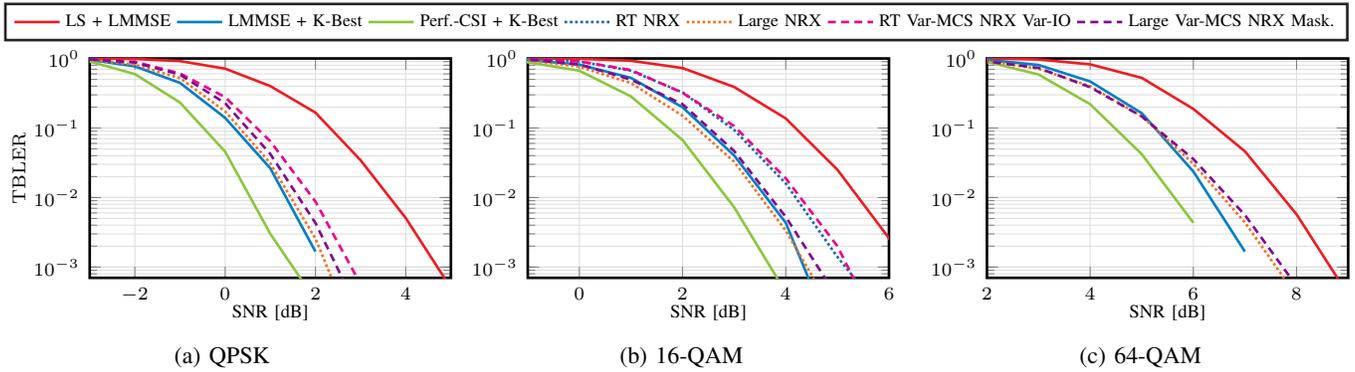

\subsection{Training with Mixed Modulation and Coding Schemes}\label{sec:var_mcs_training}
It has been empirically observed that training a \gls{Var-MCS-NRX} requires additional considerations beyond those mentioned in \cref{sec:preliminary_training}, which we detail in the following:

\subsubsection{Random MCS-to-UE association}
\comsc{}
For each training sample and each \gls{UE}, we sample an \gls{iid} random \gls{MCS} index from a set of supported \gls{MCS}. This ensures that both, single-\gls{MCS} and mixed-\gls{MCS} transmission scenarios, are represented in each training batch.
Note that explicitly training for all possible \gls{MCS}-to-\gls{UE} associations is infeasible due to the large number of different transmission scenarios. E.g., for eight active \glspl{UE} and four \glspl{MCS}, we find 165 \emph{different} associations even without considering the order or cases where not all \glspl{UE} are active.

\subsubsection{MCS specific SNR offsets}
As higher \gls{MCS} and a larger number of active users typically result in higher error-rates and higher training loss for a given noise variance $N_0$, we apply offsets to the random training \gls{SNR} (in decibels) of each batch sample depending on the random number of active \glspl{UE} and depending on the random \gls{MCS}-to-\gls{UE} association. This ensures that batch samples with higher \gls{MCS} indices are trained (on average) on larger \gls{SNR} values than batch samples with lower \gls{MCS} indices. Thereby, we can avoid that the batch loss is dominated by batch samples with many high-\gls{MCS} \glspl{UE}.

\subsection{Simulation Results}\label{sec:var_mcs_results}
We now evaluate the \gls{TBLER} performance of the \gls{Var-MCS-NRX} in a \gls{MUMIMO} scenario with $U=2$ \glspl{UE} transmitting in \gls{UL} direction to a \gls{BS} with $B=4$ antennas. The \gls{BS} has dual-polarized antennas with 3GPP TR 38.901 antenna patterns,
arranged in a horizontal \gls{ULA}. The \glspl{UE} implement two single-polarized omnidirectional antennas which are also arranged in a horizontal \gls{ULA}, and apply beamforming with $\vv=[1,1]^T/\sqrt{2}$.
In this section, we adopt the 3GPP \gls{UMi} channel model for training.
For evaluation, we combine two \gls{TDL} models, namely 3GPP TDL-B with $400$\,Hz Doppler spread and $100$\,ns delay spread for the first user and 3GPP TDL-C with $100$\,Hz Doppler spread and $300$\,ns delay spread for the second user, respectively. In the following, we denote the resulting TDL channel model as \emph{DoubleTDL} channel.
We simulate 5G NR compliant \gls{OFDM} slots with a carrier frequency of $2.14$\,GHz and a bandwidth of approximately $47.5$\,MHz, which equals 132 \glspl{PRB}, each of which consists of 12 subcarriers spaced with $30$\,kHz.
For the pilots, we select \emph{\gls{DMRS} type A} with one additional \gls{DMRS} position.

We compare two different \gls{NRX} architectures both with a feature depth of $d_{\textnormal{S}}=56$. We denote the first architecture as \emph{Large NRX} which consist of $N_{\textnormal{it}}=8$ \gls{CGNN} iterations and a total of $4.4\cdot10^5$ weights. A reduced architecture is denoted as \emph{Real-time (RT) NRX} and consists of only $N_{\textnormal{it}}=2$ \gls{CGNN} iterations resulting in only $1.4\cdot10^5$ weights.
The \gls{Var-MCS-NRX} stores $0.4\cdot10^5$ additional weights per set of IO layers.

\cref{fig:variable_mcs_big_system} compares the \gls{TBLER} performance of various \glspl{NRX} and classical baselines. The RT \gls{Var-MCS-NRX} is trained by randomly sampling \gls{iid} uniformly from \gls{MCS} index $i=9$ (QPSK) and $i=14$ (16-QAM), and the Large \gls{Var-MCS-NRX} is trained by additionally sampling \gls{iid} uniformly from $i=19$ (64-QAM).\footnote{For 64-QAM, we observed a larger performance gap between the RT and the Large \gls{NRX} architecture as compared to QPSK and 16-QAM. Hence, we consider the RT \gls{NRX} architecture only for detecting QPSK and 16-QAM.}
Note that the selection of $i\in\{9,14,19\}$ covers \emph{all} modulation orders defined in \cite[Table 5.1.3.1-1]{38214}.
The results in \cref{fig:variable_mcs_big_system} show that all \gls{NRX} architectures are capable of approaching the performance of the \gls{LMMSE} channel estimation baseline with K-Best detection.\footnote{Covariance matrices for \gls{LMMSE} channel estimation are computed from the 3GPP \gls{UMi} model and the K-Best detector applies a list size of $k=64$.} In all scenarios, the \glspl{Var-MCS-NRX} implementations closely approach the performance of their single-\gls{MCS} \gls{NRX} counterparts.

\usetikzlibrary{shapes.geometric, patterns.meta, arrows, calc}

\section{NRX Complexity and Real-time Architecture}\label{sec:rt_nrx}

Practical deployment of the \gls{NRX} requires real-time inference capabilities, which imposes strict constraints on the computational latency of the underlying \gls{NN} architecture. Further, the real-time aspects prevent from processing multiple samples in parallel (\emph{inter-frame} parallelization) and only \emph{intra-frame} parallelization can be utilized as buffering would be required otherwise.
We assume a strict computational latency budget of $1$\,ms for the \gls{NRX} using an NVIDIA A100 GPU. Furthermore, we assume inline acceleration, i.e., we ignore any \emph{memcopy} latencies from host to device and vice versa.

\subsection{Latency Measurements \& Optimization}

As mentioned earlier, it is a non-trivial task to predict the inference latency only from the model description. This stems from the fact that during inference many processing steps happen in parallel and also the memory access can become the bottleneck. To get a more realistic latency measure, we deploy the trained \gls{NRX} model using TensorRT as real-time inference engine. The resulting TensorRT engine applies advanced optimization techniques tailored to the targeted inference hardware (and software stack) such as the fusing of operations during model inference. This process resembles the compilation of source code to a target deployment platform.

Weights are quantized to \emph{float16}. However, we did not implement quantization-aware training techniques in the scope of this work. Exploring even lower quantization levels such as \emph{float8} or \emph{int8} precision is a subject of future research.

Based on the detailed profiling output of the TensorRT deployment, we carefully adjust the TensorFlow model. Removing compute (and memory) bottlenecks is a cumbersome task that requires carefully performed optimization steps. For brevity, the details are omitted here (though the optimized architecture is available in the code release). We like to emphasize that even if a bottleneck is identified, removing the operation may not solve the problem, as one also needs to understand its impact on the \gls{SNR} performance of the \gls{NRX}.

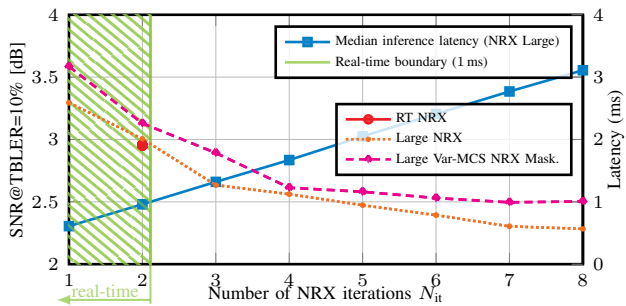
\begin{figure}[t]
    \centering
    \begin{tikzpicture}
        \pgfplotsset{compat=1.5}
        \tikzset{font={\fontsize{7pt}{7}\selectfont}}
        \begin{axis}[
            xmode=normal,
            ymode=normal,
            ylabel={\gls{SNR}@\gls{TBLER}=10\% [dB]},
            xlabel=Number of NRX iterations $N_{\textnormal{it}}$,
            y label style={at={(axis description cs:-0.065,.5)},anchor=south},
            x label style={at={(axis description cs:0.5,-0.05)},anchor=north},
            xmin = 1,
            xmax = 8,
            ymax = 4,
            ymin = 2,
            mark size=1.5pt,
            legend style={nodes={scale=0.75, transform shape}},
            legend style={at={(0.98,0.5)},anchor=east},
            grid=both,
            minor grid style={gray!25},
            width=.95\columnwidth,
            height=4.9cm,
            legend cell align={left},
            line width=1pt,
            cycle list name=corporate_colours_markers]
            \fill [opacity=.75, pattern={Lines[angle=140,yshift=-1.5pt]}, pattern color=apfelgruen] (axis cs: 2.11,0) -- (axis cs: 2.11,4) -- (axis cs: 0,4) -- (axis cs: 0,0) -- cycle;
            \draw[color=apfelgruen, opacity=.75] (axis cs: 2.11, 0) -- (axis cs: 2.11, 4);
            \addplot table[x expr=2, y=snr_at_bler_nrx_rt, col sep=comma]{./results/csv/2024-07-20_17-44_snr_at_bler_vs_nrx_iter_latency_tdl.csv};
            \pgfplotsset{cycle list shift=3}
            \addplot table[x=num_nrx_iter, y=snr_at_bler_nrx_large, col sep=comma]{./results/csv/2024-07-20_17-44_snr_at_bler_vs_nrx_iter_latency_tdl.csv};
            \addplot table[x=num_nrx_iter, y=snr_at_bler_nrx_large_masking, col sep=comma]{./results/csv/2024-07-20_17-44_snr_at_bler_vs_nrx_iter_latency_tdl.csv};
            \coordinate (p1) at (axis cs: 2.11,4);
        \end{axis}

        \begin{axis}[
            axis y line*=right,
            axis x line=none,
            ymin=0.5, ymax=6,
            ylabel={Latency (ms)},
            xmode=normal,
            ymode=normal,
            y label style={at={(axis description cs:1.1,.5)},anchor=south},
            x label style={at={(axis description cs:0.5,-0.05)},anchor=north},
            xmin = 1,
            xmax = 8,
            ymax = 4,
            ymin = 0,
            mark size=1.5pt,
            legend style={nodes={scale=0.75, transform shape}},
            legend style={at={(.98, 0.85)},anchor=east, fill=white, fill opacity=0.8, draw opacity=1,text opacity=1},
            grid=none,
            minor grid style={gray!25, densely dotted},
            width=.95\columnwidth,
            height=4.9cm,
            legend cell align={left},
            line width=1pt,
            cycle list name=corporate_colours_markers
          ]
          \pgfplotsset{cycle list shift=1}
          \addplot table[x=num_nrx_iter, y=median_latency_a_100, col sep=comma]{./results/csv/2024-07-20_17-44_snr_at_bler_vs_nrx_iter_latency_tdl.csv};
          \addlegendentry{Median inference latency (NRX Large)}

          \addlegendimage{color=apfelgruen, opacity=.75}
          \addlegendentry{Real-time boundary ($1$\,ms)}
        \end{axis}

        \begin{axis}[
            xmode=normal,
            ymode=normal,
            ylabel={\gls{SNR}@\gls{TBLER}=10\% [dB]},
            xlabel=Number of NRX iterations $N_{\textnormal{it}}$,
            y label style={at={(axis description cs:-0.065,.5)},anchor=south},
            x label style={at={(axis description cs:0.5,-0.05)},anchor=north},
            xmin = 1,
            xmax = 8,
            ymax = 4,
            ymin = 2,
            mark size=1.5pt,
            legend style={nodes={scale=0.75, transform shape}},
            legend style={at={(0.98,0.5)},anchor=east, fill=white, fill opacity=0.8, draw opacity=1,text opacity=1},
            grid=none,
            minor grid style={gray!25},
            width=.95\columnwidth,
            height=4.9cm,
            legend cell align={left},
            line width=1pt,
            axis lines = none,
            cycle list name=corporate_colours_markers]

            \fill [opacity=.75, pattern={Lines[angle=140,yshift=-1.5pt]}, pattern color=apfelgruen] (axis cs: 2.11,0) -- (axis cs: 2.11,4) -- (axis cs: 0,4) -- (axis cs: 0,0) -- cycle;
            \draw[color=apfelgruen, opacity=.75] (axis cs: 2.11, 0) -- (axis cs: 2.11, 4);

            \addplot table[x expr=2, y=snr_at_bler_nrx_rt, col sep=comma]{./results/csv/2024-07-20_17-44_snr_at_bler_vs_nrx_iter_latency_tdl.csv};
            \addlegendentry{RT NRX}
            \pgfplotsset{cycle list shift=3}
            \addplot table[x=num_nrx_iter, y=snr_at_bler_nrx_large, col sep=comma]{./results/csv/2024-07-20_17-44_snr_at_bler_vs_nrx_iter_latency_tdl.csv};
            \addlegendentry{Large NRX}
            \addplot table[x=num_nrx_iter, y=snr_at_bler_nrx_large_masking, col sep=comma]{./results/csv/2024-07-20_17-44_snr_at_bler_vs_nrx_iter_latency_tdl.csv};
            \addlegendentry{Large Var-MCS NRX Mask.}

            \coordinate (p1) at (axis cs: 2.11,2);
        \end{axis}
        
        \draw[thick, color=apfelgruen, opacity=.7] (p1) to ++(0,-0.525);
        \draw[-latex, thick, color=apfelgruen, opacity=.7] ($(p1)-(0,.475)$) -- node[above left=-0.1cm and -0.6cm] {\scriptsize real-time} ++ (-1.25,0);
    \end{tikzpicture}
    \caption{\gls{SNR} performance vs. number of NRX iterations for $U\times B \equiv 2\times4$ \gls{MUMIMO}, Double-TDL channels and 16-QAM. NRX inference latency measured on NVIDIA A100.}
    \label{fig:performance_vs_complexity_nrx_2x4}
\end{figure}

\subsection{Controlling the Inference Latency}

As mentioned in Sec.~\ref{sec:preliminary_training}, we incorporate a multi-loss \cite{nachmani2016learning} in the \gls{NRX} training pipeline which enables to adjusting the depth of the \gls{NRX} during inference without the need for any re-training. Thereby, we can control the receiver's computational latency and the accuracy of the \gls{NN} after training, e.g, to adapt to new hardware platforms or varying system configurations.

Fig.~\ref{fig:performance_vs_complexity_nrx_2x4} shows the required \gls{SNR} to achieve a target \gls{TBLER} of 10\% evaluated for different receiver depths $N_{\textnormal{it}}\in\{1,\dots,8\}$. The Large NRX is trained only once for $N_\textnormal{it}=8$ \gls{CGNN} iterations (using multi-loss).
The second axis in Fig.~\ref{fig:performance_vs_complexity_nrx_2x4} shows the corresponding latency in milliseconds evaluated on an NVIDIA A100 GPU. The latency of the \gls{NRX} is measured using the exported TensorRT engine and increases linearly with the number of iterations. This is to be expected, as each iteration has the same cost. For the given system configuration of 132 PRBs and 2 active \glspl{UE}, each iteration requires approximately $350$\,µs and we observe a constant initialization (and readout) overhead of $270$\,µs.

As can be seen in Fig.~\ref{fig:performance_vs_complexity_nrx_2x4}, the strict computational latency constraint of $1$\,ms restricts the receiver to $N_{\textnormal{it}}=2$. For comparison, we also evaluate another \gls{NRX} architecture that was trained for 2 iterations only (\emph{RT NRX}). The performance degradation of the adaptive receiver version is almost negligible.
From the \gls{TBLER} curves, it follows that $N_{\textnormal{it}}=2$ is a sub-optimal solution for the achievable error-rate performance. This implies that in a practical deployment scenario, the achievable \gls{TBLER} performance of the proposed \gls{NRX} is limited by its inference latency (and computational complexity).

The proposed real-time NRX architecture is applicable for practical deployment and achieves a strong performance when compared to classical baselines. However, the remaining gap to the larger NRX shows that there is still some potential for future work.
Finally, we want to underline that these experiments including the training results are available online.$^1$


\vspace*{-0.2cm}
\section{Site-specific Fine-tuning}
In this section, we investigate how well a generalized \gls{NRX} performs after site-specific deployment in a specific radio environment. Furthermore, we want to answer how much training data is needed for site-specific fine-tuning and how many fine-tuning training steps are required to improve the performance over the pre-trained generalized \gls{NRX}.

\begin{figure}[t]
    \centering
    \includegraphics[width=.7\columnwidth]{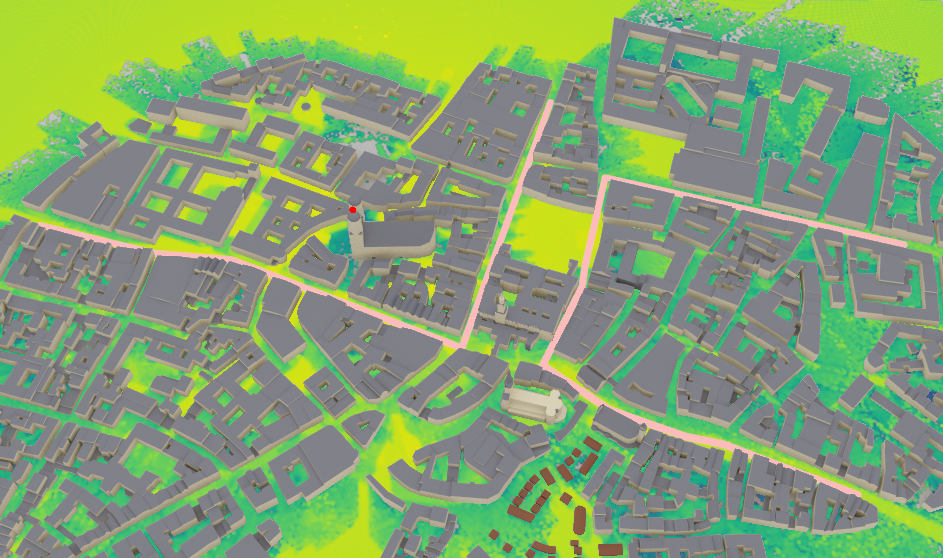}
    \caption{Ray tracing environment using Sionna's Munich map, augmented with coverage map, visualizing \gls{BS} as red point, and \gls{UE} evaluation trajectories as salmon-colored lines.}
    \label{fig:rtx_munich_map}
\end{figure}

\subsection{Ray Tracing-based Training and Evaluation Dataset}

We deploy the \gls{BS} antenna array in Sionna's ray tracer \cite{hoydis2023sionna} in the Munich map on top of a church tower of the Frauenkirche, as depicted by the red dot in \cref{fig:rtx_munich_map}. We equip the \gls{BS} and \glspl{UE} with the same antenna configurations as in the experiments in \cref{sec:rt_nrx}.
However, as the \gls{BS} should cover the whole area all around the church tower, we apply isotropic antenna characteristics instead of the 3GPP TR 38.901 antenna pattern used for \gls{NRX} pre-training.

For generating the training dataset, we randomly sample positions from the coverage map and compare different numbers of training data samples $N_{\textnormal{TD}}$. We only add data samples that have at least one valid path between any transmit and receive antenna, for all of the $T=14$ \gls{OFDM} symbols. Together with each random position, we also sample random \gls{UE} velocities uniformly in $[0, 8]$\,m/s, independently for both, the x- and y-velocity, which induces a Doppler shift to the \gls{CIR}.

The evaluation dataset is generated from two trajectories (one for each \gls{UE}, depicted by the salmon-colored lines in \cref{fig:rtx_munich_map}), whereof we sample $10^4$ positions uniformly between the starting and end points. Each of the two \glspl{UE} moves at a constant speed, $3.5$\,m/s and $3.0$\,m/s, respectively. Sample indices where the ray-tracer did not find any valid path for either of the two \glspl{UE} are removed. During simulation, we construct \gls{MUMIMO} channels by randomly sampling \emph{two different} indices in $\{1,\dots,10^4\}$ for each batch sample, which selects one position for each \gls{UE} from their respective trajectory (random sub-sampling).

\subsection{Effect of dataset Size and Fine-tuning Training Iterations}

\begin{figure}[t]
    \centering
    \begin{tikzpicture}
        \pgfplotsset{compat=1.5}
        \tikzset{font={\fontsize{7pt}{7}\selectfont}}
        \begin{axis}[
            xmode=log,
            ymode=normal,
            ylabel={SNR@TBLER=10\% [dB]},
            xlabel=Number of fine-tuning iterations $N_{\textnormal{FT}}$,
            y label style={at={(axis description cs:-0.065,.5)},anchor=south},
            x label style={at={(axis description cs:0.5,-0.075)},anchor=north},
            xmin = 1e2,
            xmax = 1e5,
            ymax = 12,
            ymin = 7.5,
            mark size=1.5pt,
            ytick={8,9,10,11,12,13},
            legend style={nodes={scale=0.65, transform shape}},
            legend pos = north east,
            grid=both,
            minor grid style={gray!25},
            major grid style={gray!25},
            width=.95\columnwidth,
            height=4.9cm,
            legend cell align={left},
            line width=1pt,
            cycle list name=corporate_colours_no_markers1]
            \addplot+[mark=none, line width=1pt] table[x=num_ft_iter, y=nrx_baseline, col sep=comma]{./results/csv/2024-07-25_08-37_site_specific_snr_at_bler_vs_num_ft_iter.csv};
            \addlegendentry{Pre-trained (UMi)}
            \label{pgf:p1}
            \addplot table[x=num_ft_iter, y=nrx_100samples, col sep=comma]{./results/csv/2024-07-25_08-37_site_specific_snr_at_bler_vs_num_ft_iter.csv};
            \addlegendentry{$N_{\textnormal{TD}}=100$}
            \addplot table[x=num_ft_iter, y=nrx_200samples, col sep=comma]{./results/csv/2024-07-25_08-37_site_specific_snr_at_bler_vs_num_ft_iter.csv};
            \addlegendentry{$N_{\textnormal{TD}}=200$}
            \addplot table[x=num_ft_iter, y=nrx_1000samples, col sep=comma]{./results/csv/2024-07-25_08-37_site_specific_snr_at_bler_vs_num_ft_iter.csv};
            \addlegendentry{$N_{\textnormal{TD}}=10^3$}
            \addplot table[x=num_ft_iter, y=nrx_10000samples, col sep=comma]{./results/csv/2024-07-25_08-37_site_specific_snr_at_bler_vs_num_ft_iter.csv};
            \addlegendentry{$N_{\textnormal{TD}}=10^4$}

            \pgfplotsset{cycle list shift=-5}
            \addplot+[densely dotted, mark=none, line width=1pt] table[x=num_ft_iter, y=nrx_baseline_umi, col sep=comma]{./results/csv/2024-07-25_08-37_site_specific_snr_at_bler_vs_num_ft_iter.csv};
            \label{pgf:p2}
            \addplot+[densely dotted] table[x=num_ft_iter, y=nrx_100samples_umi, col sep=comma]{./results/csv/2024-07-25_08-37_site_specific_snr_at_bler_vs_num_ft_iter.csv};
            \addplot+[densely dotted] table[x=num_ft_iter, y=nrx_200samples_umi, col sep=comma]{./results/csv/2024-07-25_08-37_site_specific_snr_at_bler_vs_num_ft_iter.csv};
            \addplot+[densely dotted] table[x=num_ft_iter, y=nrx_1000samples_umi, col sep=comma]{./results/csv/2024-07-25_08-37_site_specific_snr_at_bler_vs_num_ft_iter.csv};
            \addplot+[densely dotted] table[x=num_ft_iter, y=nrx_10000samples_umi, col sep=comma]{./results/csv/2024-07-25_08-37_site_specific_snr_at_bler_vs_num_ft_iter.csv};

            \node [draw, fill=white] at (rel axis  cs: 0.58, .27) {\fontsize{6pt}{6}\selectfont\shortstack[l]{
            \ref{pgf:p1} site-specific\\
            \ref{pgf:p2} 3GPP UMi
            }};
        \end{axis}
    \end{tikzpicture}
    \vspace*{-0.2cm}
    \caption{SNR performance of RT NRX with fine-tuned weights for varying $N_{\textnormal{FT}}$ and $N_{\textnormal{TD}}$ in $U\times B \equiv 2\times4$ \gls{MUMIMO} with \gls{MCS} index $i=14$. Solid curves benchmarked on site-specific evaluation dataset, dashed curves on 3GPP \gls{UMi} channels.}
    \label{fig:site_specific_sweep}
\end{figure}
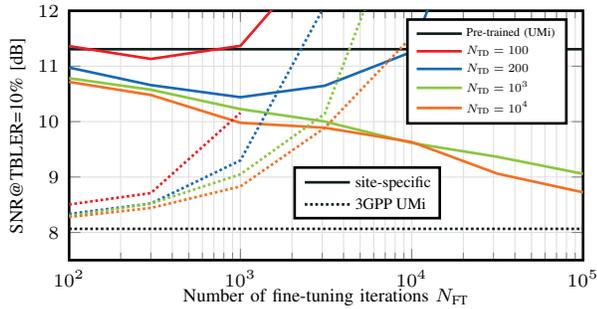

We now compare the effect of the training dataset size $N_{\textnormal{TD}}$ and the number of fine-tuning training iterations $N_{\textnormal{FT}}$. For fine-tuning training, we start with the \gls{NRX} weights from \cref{sec:rt_nrx} that have been extensively pre-trained on the 3GPP \gls{UMi} channel model. Then, we apply $N_{\textnormal{FT}}$ gradient-descent steps with the site-specific training data. Since ground-truth \gls{CSI} data is typically not available in real-world datasets, we fine-tune without double-readout.

In \cref{fig:site_specific_sweep}, we visualize the performance gains achieved with site-specific training. On the y-axis we compare the required SNR to achieve a \gls{TBLER} of 10\% for different training dataset sizes $N_{\textnormal{TD}}$. The solid curves are evaluated on the ray tracing-based evaluation dataset. We can see that very small datasets (with $N_{\textnormal{TD}}\in\{100, 200\}$) only lead to improvements in the early fine-tuning iterations, but show degraded performance for $N_{\textnormal{FT}}>10^3$. With the two larger datasets ($N_{\textnormal{TD}}\in\{10^3, 10^4\}$), the \gls{SNR} performance of the \gls{NRX} increases up to $N_{\textnormal{FT}}=10^5$.
We expect this phenomena to be caused by overfitting of the \gls{NRX} to the small datasets. Note that $N_{\textnormal{FT}}=10^3$ fine-tuning steps only take about 30\,s on an NVIDIA RTX 3090 GPU.

The dotted curves in \cref{fig:site_specific_sweep} visualize the \gls{SNR} performance of the site-specific \gls{NRX} weights evaluated on the 3GPP \gls{UMi} channel model.
These curves show the effect of \emph{catastrophic forgetting}, as we see degraded performance on \gls{UMi} channels, the better the \gls{NRX} is  fine-tuned to the ray tracing-based radio environment. Catastrophic forgetting happens more with smaller training datasets, where the effect of overfitting is more dominant. Note that more advanced training schemes based on concepts of \emph{transfer learning} may help to further improve the training convergence and to reduce the effect of catastrophic forgetting. See \cite{fischer2022adaptive1} for an early investigation.

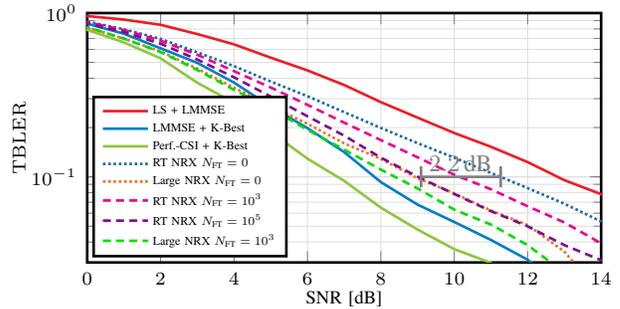
\begin{figure}[t]
    \centering
    \begin{tikzpicture}
        \pgfplotsset{compat=1.5}
        \tikzset{font={\fontsize{7pt}{7}\selectfont}}
        \begin{axis}[
            xmode=normal,
            ymode=log,
            xlabel={SNR [dB]},
            ylabel=$\mathrm{TBLER}$,
            y label style={at={(axis description cs:-0.1,.5)},anchor=south},
            x label style={at={(axis description cs:0.5,-0.075)},anchor=north},
            xmin = 0,
            xmax = 14,
            ymax = 1e-0,
            ymin = 3e-2,
            mark size=1.5pt,
            legend style={nodes={scale=0.65, transform shape}},
            legend style={at={(0.01,0.02)},anchor=south west},
            grid=both,
            minor grid style={gray!25},
            major grid style={gray!25},
            width=.95\columnwidth,
            height=4.9cm,
            legend cell align={left},
            line width=1pt,
            cycle list name=corporate_colours_no_markers]
            \addplot table[x=snr_db, y=BaselineLSlinLMMSE20, col sep=comma]{./results/csv/2024-07-23_12-52_nrx_site_specific_baseline.csv};
            \addlegendentry{LS + LMMSE}
            \addplot table[x=snr_db, y=BaselineLMMSEKBest20, col sep=comma]{./results/csv/2024-07-23_12-52_nrx_site_specific_baseline.csv};
            \addlegendentry{LMMSE + K-Best}
            \addplot table[x=snr_db, y=BaselinePerfCSIKBest20, col sep=comma]{./results/csv/2024-07-23_12-52_nrx_site_specific_baseline.csv};
            \addlegendentry{Perf.-CSI + K-Best}
            \addplot table[x=snr_db, y=NeuralReceiver20, col sep=comma]{./results/csv/2024-07-23_12-52_nrx_site_specific_baseline.csv};
            \addlegendentry{RT NRX $N_{\textnormal{FT}}=0$}
            \addplot table[x=snr_db, y=NeuralReceiver20, col sep=comma]{./results/csv/2024-07-23_12-52_nrx_site_specific_baseline_large.csv};
            \addlegendentry{Large NRX $N_{\textnormal{FT}}=0$}
            \addplot table[x=snr_db, y=NeuralReceiver20, col sep=comma]{./results/csv/2024-07-25_08-28_nrx_site_specific.csv};
            \addlegendentry{RT NRX $N_{\textnormal{FT}}=10^3$}
            \addplot table[x=snr_db, y=NeuralReceiver20, col sep=comma]{./results/csv/2024-07-25_08-28_nrx_site_specific_100k.csv};
            \addlegendentry{RT NRX $N_{\textnormal{FT}}=10^5$}
            \addplot+[densely dashed] table[x=snr_db, y=NeuralReceiver20, col sep=comma]{./results/csv/2024-07-23_12-52_nrx_site_specific_large.csv};
            \addlegendentry{Large NRX $N_{\textnormal{FT}}=10^3$}

            \draw[|-|,gray] (axis cs: 9.059, 1e-1) -- (axis cs: 11.301, 1e-1) node[above left=-.065cm and .0cm] {\small $2.2$\,dB};

        \end{axis}
    \end{tikzpicture}
    \vspace*{-0.2cm}
    \caption{\gls{TBLER} vs. \gls{SNR} with site-specific evaluation dataset in $U\times B \equiv 2\times4$ \gls{MUMIMO} with \gls{MCS} index $i=14$. Fine-tuned \glspl{NRX} trained with $N_{\textnormal{TD}}=10^3$ samples.}
    \label{fig:site_specific_bler_vs_snr}
\end{figure}

In \cref{fig:site_specific_bler_vs_snr}, we evaluate the \gls{TBLER} of the \gls{NRX} and other \emph{classical} baselines using site-specific data. We observe that the fine-tuned RT \gls{NRX} with $N_{\textnormal{FT}}=10^5$ closely approaches the performance of the Large \gls{NRX} without fine-tuning. The fine-tuned Large \gls{NRX} with $N_{\textnormal{FT}}=10^3$ closely approaches the K-Best baseline with \gls{LMMSE} channel estimation.

\vspace*{-0.2cm}
\section{Conclusion}
\vspace*{-0.1cm}

We have proposed solutions for deploying of \glspl{NRX} in real-world 5G NR systems: (i) adaptability for operation with variable and mixed \gls{MCS} conditions, (ii) a real-time architecture that meets the latency requirements of practical software-defined 5G NR systems, and (iii) generalized \gls{NRX} pre-training with significant performance gains through site-specific fine-tuning, requiring only a few thousand iterations and data samples.
We leaved fine-tuning on real-world measurement datasets as well as a system-level performance evaluation for future work.

\end{document}